\documentclass[%
reprint,
nofootinbib,
 amsmath,amssymb,
 aps,
]
{revtex4-1}
\usepackage{epsfig,epsf,rotating,wrapfig,tabularx}
\usepackage{dcolumn}
\usepackage{bm}
\usepackage{graphicx}

\begin{document}

\preprint{APS/123-QED}

\title{Role of quarks in hadroproduction in high energy collisions}

\author{\firstname{A.~A.}~\surname{Bylinkin}}
 \email{alexandr.bylinkin@cern.ch}
\affiliation{%
 Institute for Theoretical and Experimental
Physics, ITEP, Moscow, Russia
}%
\author{\firstname{A.~A.}~\surname{Rostovtsev}}
 \email{rostov@itep.ru}
\affiliation{%
 Institute for Theoretical and Experimental
Physics, ITEP, Moscow, Russia
}%


\begin{abstract}
Qualitative model for hadroproduction in high energy collisions considering two components (``thermal''and ``hard'') to hadroproduction is proposed. Inclusive pseudorapidity distributions, $d\sigma/d\eta$, and transverse momentum spectra, $d^2\sigma/(d\eta dp_T^2)$,  measured by different collaborations are considered in terms of this model. The shapes of the pseudorapidity distributions agree with that one can expect from the qualitative picture introduced. 
 Finally, the differences between charged particle spectra produced in inclusive and diffractive events are discussed and the absence of the ``thermal'' component in the latter is observed. 
\end{abstract}

\pacs{Valid PACS appear here}
\maketitle

\section{Introduction}
\begin{figure}[h]
\includegraphics[width =8cm]{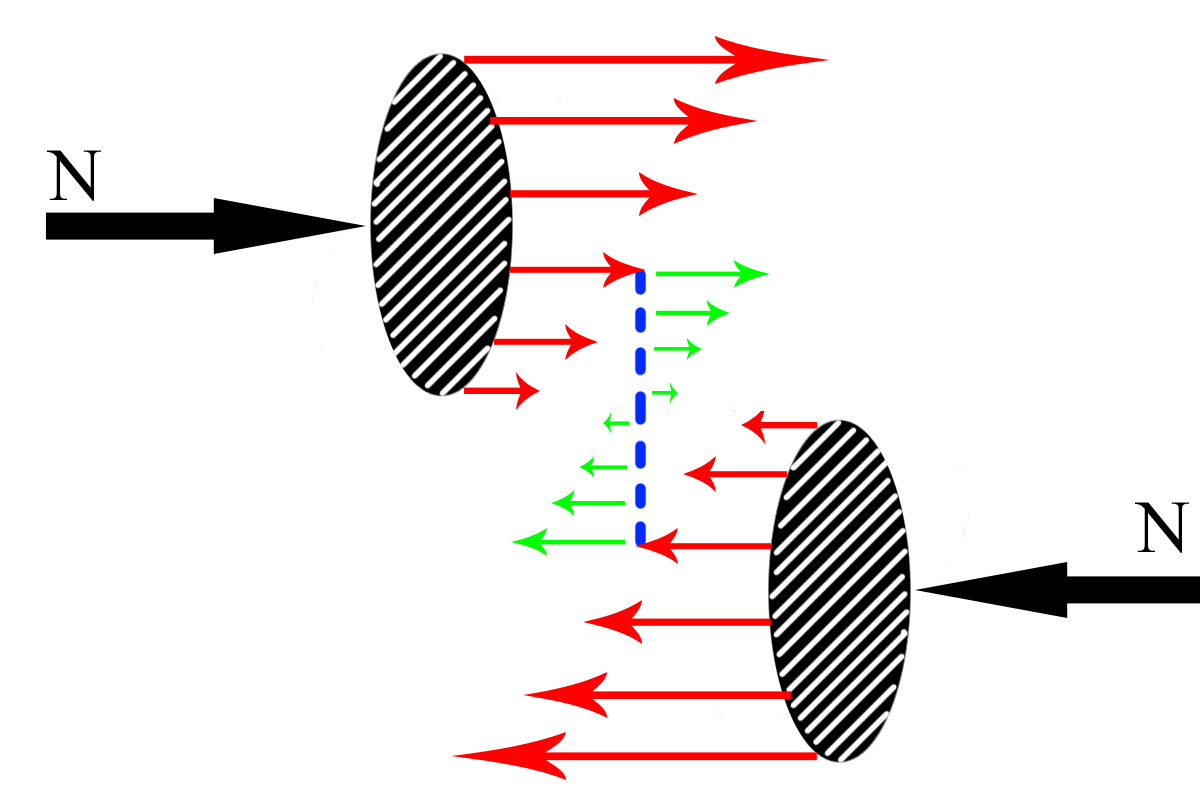}
\caption{\label{fig} Two different sources of hadroproduction: red arrows - particles produced by the preexisted partons, green - particles produced via the mini-jet fragmentation.}
\end{figure}

The baryon-baryon high energy interactions one could decompose into at least two distinct sources of produced hadrons. The first one is associated with the baryon valence quarks and a quark-gluon cloud coupled to the valence quarks. Those partons preexist long time before the interaction and could be considered as being a thermalized statistical ensemble. When a coherence of these partonic systems is destroyed via strong interaction between the two colliding baryons, these partons hadronize into particles released from the collision. The hadrons from this source are distributed presumably according to the Boltzmann-like exponential statistical distribution in transverse plane w.r.t. the interaction axis. The second source of hadrons is directly related to the mini-jet fragmentation of the virtual partons (pomeron in pQCD) exchanged between two colliding partonic systems. The radiated partons from this pomeron have presumably a typical for the pQCD power-law spectrum. Schematically figure~\ref{fig}  shows these two sources of particles produced in high energy baryonic collisions. 

\begin{figure}[h]
\includegraphics[width =8cm]{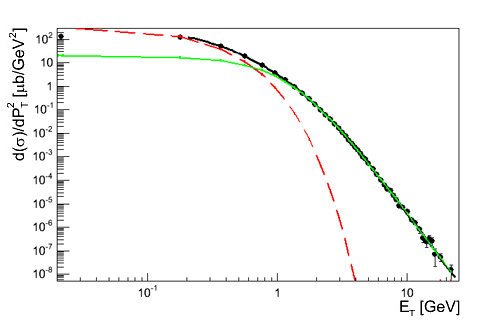}
\caption{\label{fig.0} Charge particle differential cross section~\cite{UA1} fitted to the function~(\ref{eq:exppl}): the red (dashed) line shows the exponential term and the green (solid) one - the power law.}
\end{figure}

Thus, one can study charged particle production using the {\em two component} parameteristion~\cite{OUR1}, combining an exponential (Boltzmann-like) and a power-law $p_T$ distributions:
\begin{equation}
\label{eq:exppl}
\frac{d\sigma}{p_T d p_T} = A_e\exp {(-E_{Tkin}/T_e)} +
\frac{A}{(1+\frac{p_T^2}{T^{2}\cdot n})^n},
\end{equation}
where  $E_{Tkin} = \sqrt{p_T^2 + M^2} - M$
with M equal to the produced hadron mass. $A_e, A, T_e, T, n$ are the free parameters to be determined by fit to the data.  The detailed arguments for this particular choice are given in~\cite{OUR1}.  
A typical charged particle spectrum as function of transverse energy, fitted with this function~(\ref{eq:exppl}) is shown in figure~\ref{fig.0}. 
As one can see, the exponential term dominates the particle spectrum at low $p_T$ values.

\section{Pseudorapidity distributions}
Let us first discuss the charged particle production in $pp$-collisions as function of pseudorapidity in terms of the qualitative picture for hadroproduction introduced above.  From the naive point of view, hadrons produced via the mini-jet fragmentation should be concentrated in the central rapidity region ($\eta \sim 0$), while those coming from the proton fragmentation are expected to dominate at high values of $\eta$ due to non-zero momenta of the initial partons along the beam-axis. To check this prediction  it is possible to use already available data published by the UA1 experiment~\cite{UA1} which are presented by charged particle  spectra $d^2\sigma/(d\eta dp_T^2)$ for $pp$-collision in five pseudorapidity bins, covering the total rapidity interval $|\eta|<3.0$. 

\begin{figure}[h]
\includegraphics[width =8cm]{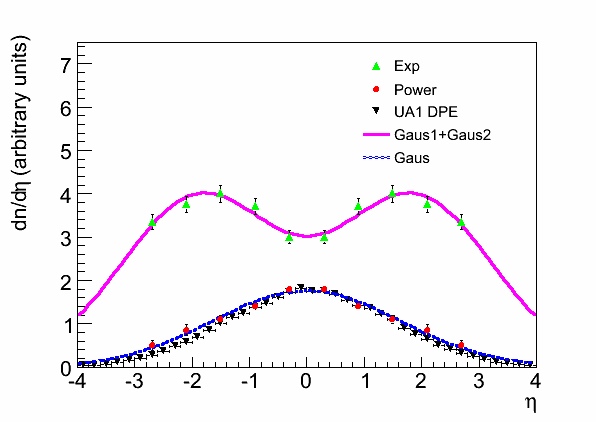}
\caption{\label{fig.1} Particle distributions calculated for power-like and exponential contributions separately and fitted with Gaussian distributions~(\ref{eq.gaus}) and (\ref{eq.gaus2}), respectively. Experimental data on double-pomeron exchange (DPE)~\cite{UA1DP} is presented with arbitrary normalization, showing good agreement in shape with the  power-like term.}
\end{figure}

The contributions $d\sigma/d\eta$ to the charged particle production from the exponential and power-like terms of~(\ref{eq:exppl}) can be studied  separately as function of $\eta$. Figure~\ref{fig.1} shows these contributions obtained from the fit (\ref{eq:exppl}) to the experimental data~\cite{UA1}. The power-like contribution is then fitted by the Gaussian distribution:
\begin{equation}
d\sigma/d\eta = A_{pl} \cdot exp[-0.5\cdot((\eta - \eta_{pl})/\sigma_{pl})^2],
\label{eq.gaus}
\end{equation}
with $\eta_{pl} = 0$, while for the exponential contribution one can assume a sum of two Gaussians:
\begin{widetext}
\begin{equation}
d\sigma/d\eta = A_{exp1} \cdot exp[-0.5\cdot((\eta - \eta_{exp1})/ \sigma_{exp1})^2]  + A_{exp2} \cdot exp[-0.5\cdot((\eta - \eta_{exp2})/\sigma_{exp2})^2],
\label{eq.gaus2}
\end{equation}
\end{widetext}
taking $A_{exp1} =  A_{exp2}$, $\sigma_{exp1} = \sigma_{exp2}$ and $\eta_{exp1} = -\eta_{exp2}$. These fits (\ref{eq.gaus}) and (\ref{eq.gaus2}) are shown in figure~\ref{fig.1} as well.

In addition, available data on the double-pomeron exchange measured at the same c.m.s. energy by the UA1 Collaboration~\cite{UA1DP} is shown in figure~\ref{fig.1}. One can notice a rather good agreement between these data~\cite{UA1DP} and the shape of the power-law term contribution obtained from the fit~(\ref{eq:exppl}), supporting the qualitative picture for hadroproduction described above. Of course, cuts on the rapidity gaps used to select the DPE events  squeeze the measured distribution, excluding events with a large $\eta$, close to the edges of the available phase space. On the other hand, particles near these edges are originated mainly from the exponential contribution. Therefore, we do not expect too much difference in the distributions for central $\eta$ corresponding to the power-like term in comparison with
the Minimum Bias (MB) events. Indeed, as it is seen in figure~\ref{fig.1}, the distribution
of the power-like component in our fit is a bit wider than that measured
by the UA1 collaboration in DPE events~\cite{UA1DP}.

\begin{figure}[h]
\includegraphics[width =8cm]{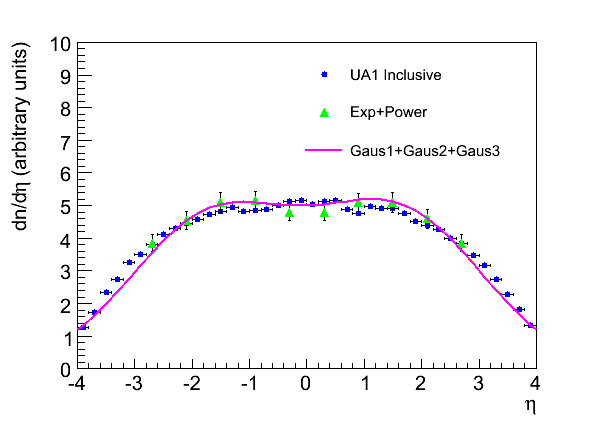}
\caption{\label{fig.2} Particle cross-sections calculated from the fit~(\ref{eq:exppl}) to the experimental data~\cite{UA1} and MB data~\cite{UA1DP} (with arbitrary normalization) shown together with a sum ((\ref{eq.gaus}) $+$ (\ref{eq.gaus2})) of three Gaussian distributions. The parameters are extracted from the fits (figure~\ref{fig.1}).}
\end{figure}

Figure~\ref{fig.2} shows the sum of (\ref{eq.gaus}) and (\ref{eq.gaus2}) together with the experimental data for MB events~\cite{UA1DP}. One can notice that the shape of the pseudorapidity distribution of charged particles is described rather well by the sum of three Gaussian distributions with the parameters extracted from the fit. Thus, the difference between the shapes of pseudorapidity distributions for DPE and MB events and the existence of a relatively wide plateau in the latter can be qualitatively explained by the introduced model.

\subsection{Scaling}
\begin{figure}[h]
\includegraphics[width =8cm]{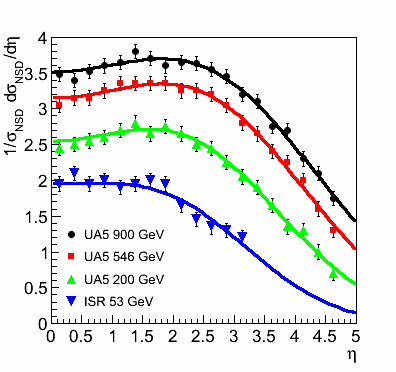}
\caption{\label{fig.3} Particle cross-sections $\frac{1}{\sigma_{NSD}} \frac{d\sigma_{NSD}}{d\eta}$~\cite{UA5, ISR} fitted with a sum ((\ref{eq.gaus}) + (\ref{eq.gaus2})) of three Gaussian distributions.}
\end{figure}

Since the shapes of pseudorapidity distributions are described by the introduced model rather well, it is interesting to study how it varies with the c.m.s. energy in a collision. This can be done, using the data on pseudorapidity distributions measured  {\em under the same experimental conditions} by the UA5 Detector~\cite{UA5,ISR} for the energies varying from $53$ to $900$ GeV. As a first step, one can extract the parameters $A_{exp}$ and $A_{pl}$ of the Gaussian distributions (\ref{eq.gaus}) and (\ref{eq.gaus2}) shown in  figure~\ref{fig.1} and extrapolate their ratio $A_{pl}/A_{exp}$ to other energies, using the dependences found recently~\cite{OURM}\footnote{Note, that in~\cite{OURM} the dependences~(\ref{eq.netap}) and (\ref{eq.netae}) are given for $\eta \sim 0$ and not for Gaussian parameters $A_{exp}$ and $A_{pl}$}:
\begin{equation}
\label{eq.netap}
(\frac{dN}{d\eta})^{power} \propto s^{0.25},
\end{equation}
\begin{equation}
\label{eq.netae}
(\frac{dN}{d\eta})^{exp} \propto s^{0.15}.
\end{equation}
Since the data in~\cite{UA5,ISR} are presented normalized to the non-single diffractive (NSD) cross-sections $\sigma_{NSD}$, one should also take into account, the growth of high energy cross-sections $\sigma_{tot}\propto s^{0.08}$~\cite{DL}, while preforming the extrapolation.  Then, one can fit the experimental data~\cite{UA5,ISR} by a sum ((\ref{eq.gaus}) $+$ (\ref{eq.gaus2})) of three Gaussian distributions. The results of this fit are shown in figure~\ref{fig.3}.

\begin{figure}[h]
\includegraphics[width =8cm]{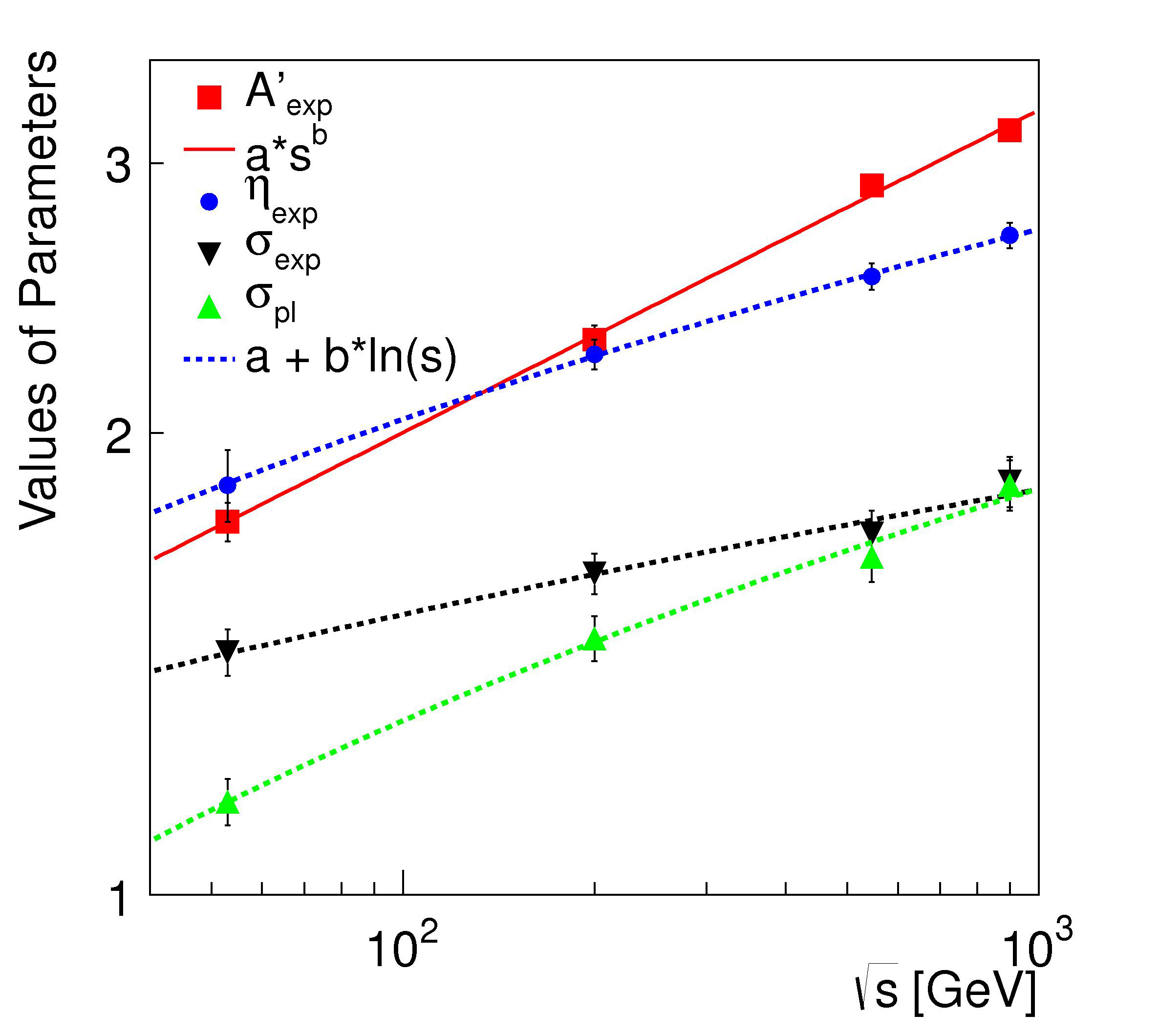}
\caption{\label{fig.4} Parameters $A'_{exp}$, $\eta_{exp}$, $\sigma_{exp}$ and $\sigma_{pl}$ of the Gaussian distributions extracted from the fit to the experimental data~\cite{UA5,ISR}. Lines show the variations of these parameters as c.m.s. energy $\sqrt{s}$.}
\end{figure}

Next, variations of the parameters of the Gaussian distributions obtained from the fit (figure~\ref{fig.3}) can be studied. The parameters $A'_{exp}$, $\eta_{exp}$, $\sigma_{exp}$ and $\sigma_{pl}$ are shown in figure ~\ref{fig.4} as function of c.m.s. energy\footnote{$A'_{exp}$ and $A'_{pl}$ correspond to charged particle densities and not to cross-sections as $A_{exp}$ and $A_{pl}$ in~(\ref{eq.gaus}) and (\ref{eq.gaus2}), repsectively}. Note, that $A'_{pl}$ can be determined from (\ref{eq.netap}) and $\eta_{pl}$ is taken to be $0$.

Finally, the variations of the parameters of the Gaussian distributions can be parametrized in the following way:
\begin{equation}
\label{eq.sigpl}
\sigma_{pl} = 0.217 + 0.235\cdot \ln\sqrt{s},
\end{equation}
\begin{equation}
\label{eq.eta}
\eta_{exp} = 0.692 + 0.293 \cdot \ln \sqrt{s},
\end{equation}
\begin{equation}
\label{eq.sigexp}
\sigma_{exp} = 0.896 + 0.136\cdot \ln \sqrt{s},
\end{equation}
\begin{equation}
\label{eq.apl}
A'_{pl} = 0.13\cdot s^{0.175},
\end{equation}
\begin{equation}
\label{eq.aexp}
A'_{exp} = 0.76\cdot s^{0.106},
\end{equation}
where $s$ is the c.m.s. energy.
\subsection{Predictions for the LHC}
\begin{figure}[!h]
\includegraphics[width =8cm]{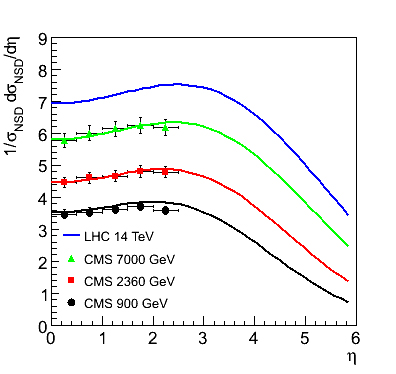}
\caption{\label{fig.5} Charged particles pseudorapidity distributions measured by the CMS Collaboration~\cite{CMS1,CMS2} and shown together with the predictions of the introduced model. Prediction for $\sqrt{s} = 14$ TeV is also shown.}
\end{figure}

These dependences~(\ref{eq.sigpl})-(\ref{eq.aexp}) can be used to make predictions on charged particles pseudorapidity distributions at LHC-energies. Such predictions can be already tested on available experimental data measured by the CMS Collaboration~\cite{CMS1,CMS2} (Figure~\ref{fig.5}).

 One can notice that the predictions made from the dependences observed (\ref{eq.sigpl})-(\ref{eq.aexp}) are in a good agreement with the experimental data up to $7$ TeV, therefore, a prediction for further LHC measurements at $14$ TeV is also shown.\\
 
\section{Charged particle production in diffractive events}
In~\cite{OUR2} it was shown that contrary to $pp$-collisions, spectra produced in $\gamma p$ or $\gamma \gamma$ collisions have no room for the exponential term. From the results shown in figure \ref{fig.1} one can also come to the similar conclusion for the charged particle spectra produced in the DPE-events in $pp$-collisions. Unfortunately, no data on the transverse momentum ($p_T$) spectra for charged particle production in such events is available at the moment. Therefore, it is suggested to look at the measurements on diffractive photoproduction ($D\gamma p$) that can be phenomenologically explained by the photon-pomeron interaction.

\begin{figure}[!h]
\includegraphics[width =8cm]{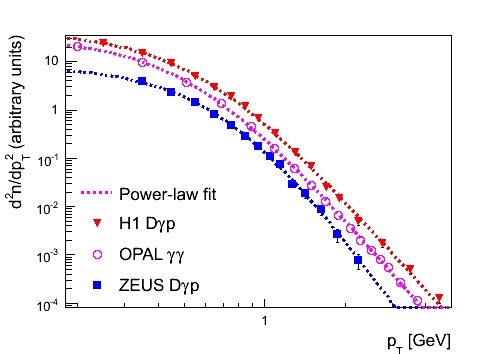}
\caption{\label{fig.6} Charged particle spectra $d^2n/dp_T^2$ measured in $D\gamma p$~\cite{ZEUS,H1D1} and $\gamma \gamma$ collisions~\cite{OPAL} fitted by the power-law term of (\ref{eq:exppl}).}
\end{figure}

Let us first consider the available data on the transverse momentum spectra produced in $D\gamma p$ events~\cite{ZEUS,H1D1}.
These spectra fitted by eq.~(\ref{eq:exppl}) are shown in the figure~\ref{fig.6}. One can notice, that similar to the case of $\gamma \gamma$ collisions (also shown in figure~\ref{fig.6}) no exponential term is needed to describe these spectra. Moreover, almost the same values of the $T$ and $N$ parameters of the power-law term in eq.~(\ref{eq:exppl}) are obtained from the fits of the $D\gamma p$ and $\gamma \gamma$ data.

\begin{figure}[!h]
\includegraphics[width =8cm]{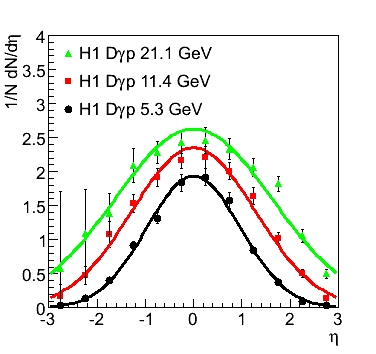}
\caption{\label{fig.7} Charged particle spectra $dn/d\eta$ measured in diffractive photoproduction $D\gamma p$~\cite{H1D2} described by only one Gaussian distribution.}
\end{figure}

Next, one can look at the pseudorapidity distributions measured in $D\gamma p$. Such distributions~\cite{H1D2} are shown together with the Gaussian fit (\ref{eq.gaus}) in figure~\ref{fig.7}. Remarkably, similar to the case of DPE in $pp$-collisions {\em only one} Gaussian form is needed to nicely describe these data. This observation further supports the hypothesis of absence of the exponential component in diffractive events.

Summarizing the observations made in this section one can conclude the following:
\begin{itemize}
\item Charged particle spectra produced in $\gamma \gamma$ and $D\gamma p$ interaction are similar in shape and both can be described by the power-law term only.
\item Pseudorapidity distributions in $D\gamma p$ have also the shape similar to those measured in DPE events and both described by {\em only one} Gaussian distribution.
\item Finally, one can conclude that the ''thermal" production expressed by the exponential term in (\ref{eq:exppl}) is essential only for $pp$-collisions and thus can be related to the presence of quarks and gluons in the initial colliding system.
\end{itemize} 

\section{Ratio between ``thermal'' and ``hard'' contributions}
\begin{figure}[!h]
\includegraphics[width =8cm]{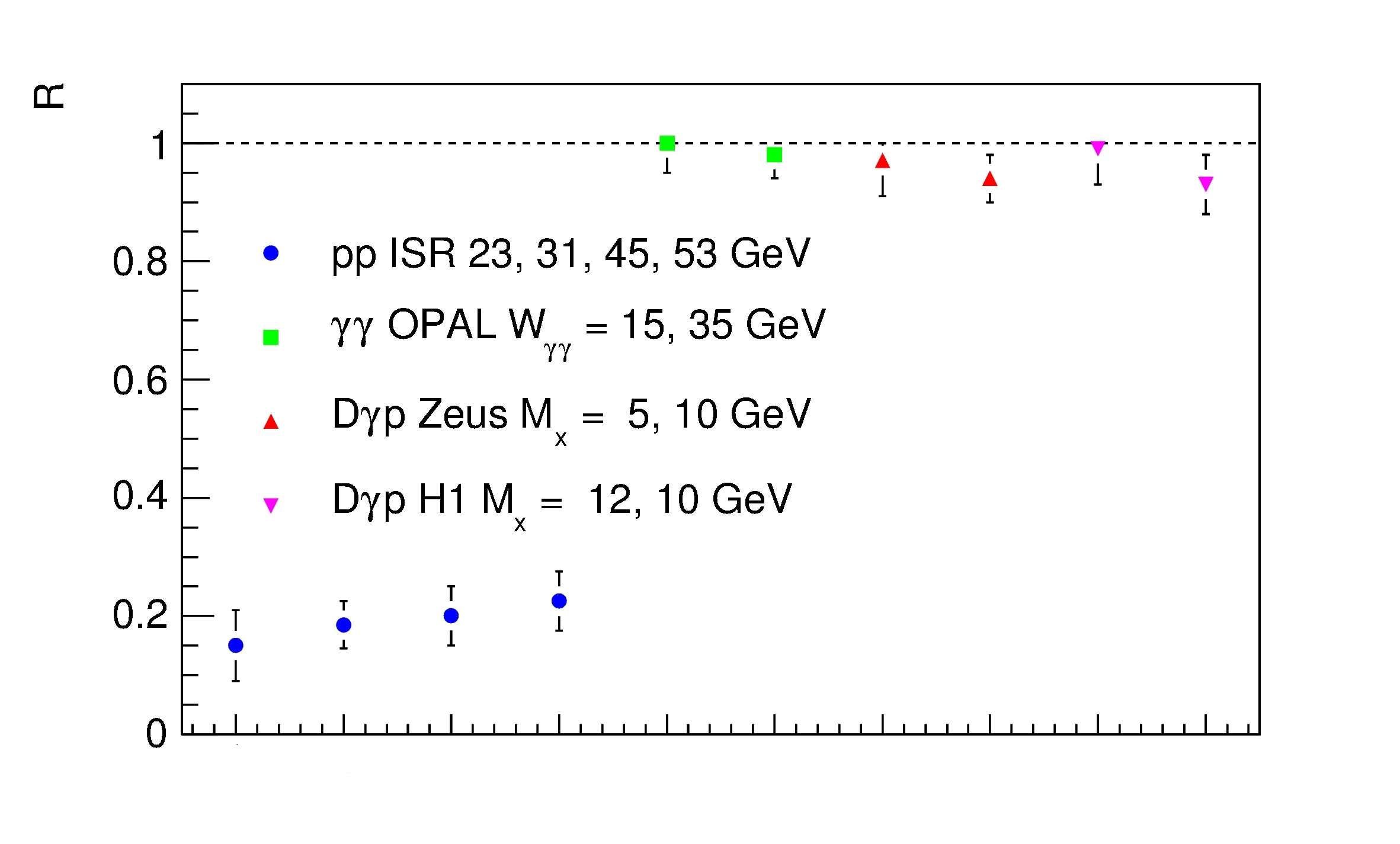}
\caption{\label{fig.7a} Value of $R$ shown for $pp$, $\gamma \gamma$ and $D\gamma p$ interactions, as calculated from the fits (\ref{eq:exppl}) to various experimental data~\cite{ISR,ZEUS,H1D1,H1D}.}
\end{figure}

In~\cite{OUR1} it was suggested to study hadroproduction dynamics using the parameter $R$:
\begin{equation}
\label{eq.r}
R = \frac{Power}{Exp + Power},
\end{equation}
 standing for the contribution of the power-law (``hard'') term to the full spectra integrated over $p_T^2$. Thus, it is interesting to look at the values of this parameter calculated from the fits (\ref{eq:exppl}) to various experimental data. The values of $R$ are shown in figure~\ref{fig.7a} for charged particle spectra measured in $pp$, $\gamma \gamma$ and $D\gamma p$ interactions together. One can notice a striking difference between these values obtained for $pp$-collisions at ISR~\cite{ISR} from those measured in $\gamma \gamma$-interaction at OPAL~\cite{OPAL} or $D\gamma p$ at HERA~\cite{ZEUS, H1D1, H1D}\footnote{The data for $pp$ and $\gamma \gamma$ interactions are chosen to have the values of $M_x$ similar to those measured in $D\gamma p$ at HERA.}. Therefore, further evidence of absence of the ``thermal'' component in diffractive events is obtained.  
 
\begin{figure}[!h]
\includegraphics[width =8cm]{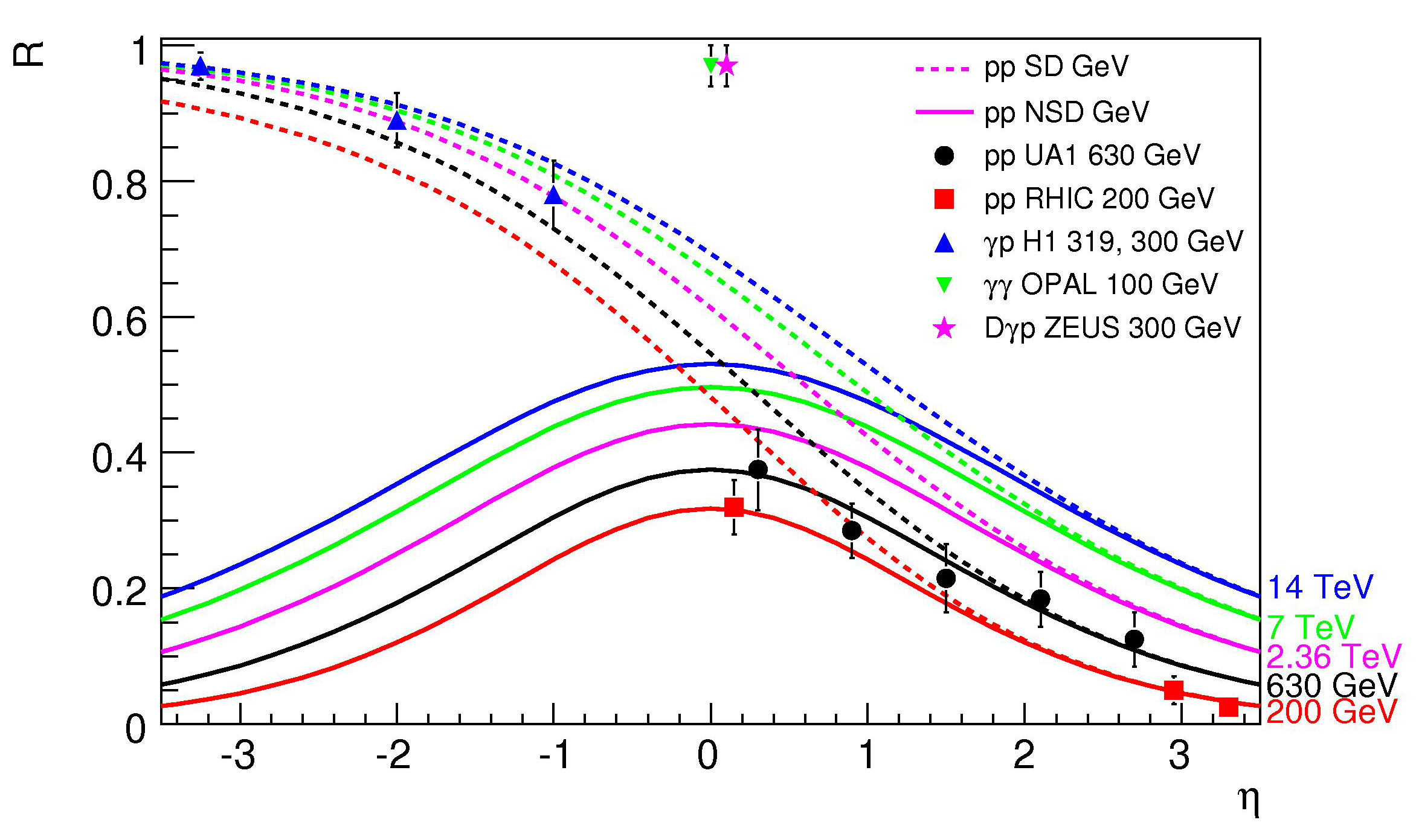}
\caption{\label{fig.8} Predictions on the contribution $R$ of the power-law term to the charged particle spectra for non-single-diffractive (NSD, solid lines) and single-diffractive (SD, dashed lines) charged particle production in $pp$ collisions. Points show the values of $R$ calculated from the fits~(\ref{eq:exppl}) to the experimental data~\cite{UA1,BRAHMS,PHENIX,OPAL,ZEUS,H11,H12}.
}
\end{figure}

  In addition, it is interesting to plot the predictions for the $R$-value, using eq. (\ref{eq.sigpl})-(\ref{eq.aexp}) and compare it with the results obtained from the fits of the transverse momentum spectra. Figure~\ref{fig.8} shows such predictions for different energies together with the fit results of PHENIX, BRAHMS and UA1~\cite{UA1,BRAHMS,PHENIX} data.

Since the similarity between $\gamma \gamma$ and $D\gamma p$ interactions has been observed one can also expect that $R$ as a function of pseudorapidity for $\gamma p$ interactions should be similar to the case of single-diffractive (SD) $pp$-collisions. Thus, predictions on $R$ for SD events and values of $R$ obtained from the fits of DIS data~\cite{H11,H12} are also shown in figure~\ref{fig.8}. One can conclude that they qualitatively agree with the behaviour predicted by the proposed model for hadroproduction.

In conclusion, qualitative model for hadroproduction in high energy collisions considering two components (``thermal'' and ``hard'') to hadroproduction has been introduced. Inclusive pseudorapidity distributions, $d\sigma/d\eta$, and transverse momentum spectra, $d^2\sigma/(d\eta dp_T^2)$,  were considered in terms of this model. The shapes of the pseudorapidity distributions agree with that one can expect from the described qualitative picture of hadroproduction. The dependences observed have been used to make predictions on the pseudorapidity distributions, $d\sigma/d\eta$, at higher energies and tested on the available experimental data. Finally, the difference between charged particle production in inclusive and diffractive processes has been discussed. Similarity between $\gamma \gamma$, $D\gamma p$ and DPE $pp$-collisions has been observed. Contrary to inclusive charged particle production in $pp$-collisions the absence of the ``thermal'' component in these processes has been observed. Thus, the ``thermal'' contribution has been related to the presence of quarks in the initial colliding system.
\begin{acknowledgements}
The authors thank Professor Mikhail Ryskin for fruitful discussions and his help provided during the preparation of this paper. 
\end{acknowledgements}

\end{document}